\documentclass[conference]{IEEEtran}
\usepackage{subfigure}
\usepackage{float}
\usepackage{cite}
\usepackage{graphicx}
\usepackage{amsmath,bm}

\begin{document}


\title{Memristive LSTM network hardware architecture for time-series predictive modeling problems}


\author{\IEEEauthorblockN{Kazybek Adam, Kamilya Smagulova and  Alex Pappachen James}
\IEEEauthorblockA{Department of Electrical and Computer Engineering\\
Nazarbayev University, 
Astana, Kazakhstan\\
Email: apj@ieee.org}}

\maketitle


\begin{abstract}
Analysis of time-series data allows to identify long term trends and make predictions that can help to improve our lives. With rapid development of artificial neural networks, long short-tern memory (LSTM) recurrent neural network (RNN) configuration is found to be capable in dealing with time-series forecasting problems where data points are time dependent and possess seasonality trends. Gated structure of LSTM cell and flexibility in network topology (one-to-many, many-to-one, etc) allows to model systems with multiple input variables and control several parameters such as the size of look-back window to make a prediction and number of time steps to be predicted. These make LSTM attractive tool over conventional methods such as  auto regression models, simple average, moving average, naive approach, ARIMA, Holt’s linear trend method, Holt’s Winter seasonal method, and others.
In this paper, we propose a hardware implementation of LSTM network architecture for time-series forecasting problem. All simulations were performed using TSMC 0.18 $\mu$m CMOS technology and HP memristor model.

\end{abstract}

\begin {IEEEkeywords}
LSTM, RNN, memristor, crossbar, analog circuit, time-series prediction
\end{IEEEkeywords}

\section{Introduction}
First introduced in 1995 \cite{hochreiter1997long}, long short-term memory (LSTM) is a special configuration of RNN aimed to bypass exploding or vanishing gradient problems in conventional RNN. This became possible due to the gated structure of an LSTM cell which allows to control the flow of current and previous cell's data.  As RNN, LSTM network  has feedback connections to retain order of information which makes it powerful tool for processing sequential data.   Various applications where LSTM networks have been successfully used include machine translation, speech recognition, forecasting, and others\cite{xingjian2015convolutional},\cite{graves2013hybrid}, \cite{soutner2013application}. Most of them are mainly software-based and few works on FPGA are introduced \cite{han2017ese}, \cite{guan2017fpga}. However, their implementation is still limited due to large complexity and parallelism of LSTM network structure that requires huge computational resources. 

\par
Previous work  \cite{smagulova2018design} offered CMOS-memristor analog circuit design of current-based LSTM cell architecture for time-series prediction problem by \cite{brownlee2016time}.  It used current mirrors and current-based activation function circuits. In this work, we propose voltage-based circuits since they provide us with higher accuracy and more predictable outputs. Current-based implementation could be used in solving a classification problem. This is because we are not interested in the analog output voltage -- as long as it is high enough or low enough, we know that it is either digital “1” or digital “0”. Whereas, in the case of time series prediction problem, we are interested in the analog output voltage to be much accurate rather than it being higher or lower of some threshold value. Therefore, high-accuracy sigmoid and hyperbolic tangent function circuits, which are voltage-based, were implemented. In addition, high accuracy four-quadrant multiplier circuit was adapted from \cite{ramirez2004low}. They help to obtain accurate values at each stage to finally arrive to an accurate output value. Additionally, control circuit has been implemented to carry out the multiple time step feature of the LSTM RNN.
\par
It is a fact that time series prediction will yield some error, for instance mean square error (MSE) or root mean square error (RMSE). If the RMSE of the circuit and the software implementations are close enough, then we can conclude that we successfully implemented the LSTM neural network in analog hardware.
\par
This paper is structured as following: first, problem description along with LSTM overview is given; further each major circuit parts are introduced; and finally simulation results are presented and conclusions are drawn.

\section{Proposed circuit}

\subsection{Problem description and LSTM overview}

Proposed circuit design implements time-series prediction problem using LSTM by Brownlee \cite{brownlee2016time}. The provided dataset demonstrates the change of the number of international airline passengers during 12 year period with 144 observation points. Prior being processed, it was divided into training and testing sets and normalized between 0 and 1 due to the sensitivity of LSTM to input data. The single-column data-set is converted into three-column, where the first column contains information on the number of passengers in the previous month, the second column in the current month and the third column is the number of passengers to be predicted (Table \ref{table1}). 
\begin{table}[ht]
\centering
\caption{The dataset for time-series forecasting problem}
\label{table1}
\begin{tabular}{cccccc}
\hline
\multicolumn{3}{c}{\textbf{Original data}}        & \multicolumn{3}{c}{\textbf{Normalized data}}      \\ \hline
\textbf{X(t-1)} & \textbf{X(t)} & \textbf{X(t+1)} & \textbf{X(t-1)} & \textbf{X(t)} & \textbf{X(t+1)} \\ \hline
112             & 118           & 132             & 0.015444        & 0.027027      & 0.054054        \\ \hline
118             & 132           & 129             & 0.027027        & 0.054054      & 0.048263        \\ \hline
132             & 129           & 121             & 0.054054        & 0.048263      & 0.032819        \\ \hline
129             & 121           & 135             & 0.048263        & 0.032819      & 0.059846        \\ \hline
...             & ...           & ...             & ...             & ...           & ...             \\ \hline
\end{tabular}
\end{table}

\begin{figure}[hb]
   \centering
   \includegraphics[scale=.3]{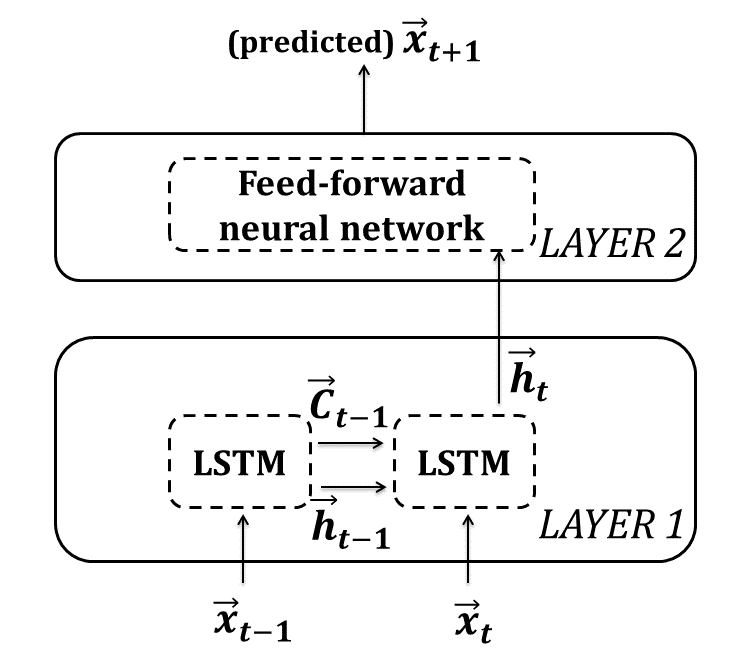}
    \caption{Implementation diagram of a time-series prediction problem using LSTM}
    \label{diagram}
\end{figure}

\par
Fig.\ref{diagram} shows a diagram for the implementation of the task. It consists of two layers, where the first layer is LSTM network with two time-steps and four hidden units and the second layer is a feed-forward neural network with linear activation function.

\begin{figure}[ht]
   \centering
   \includegraphics[scale=.27]{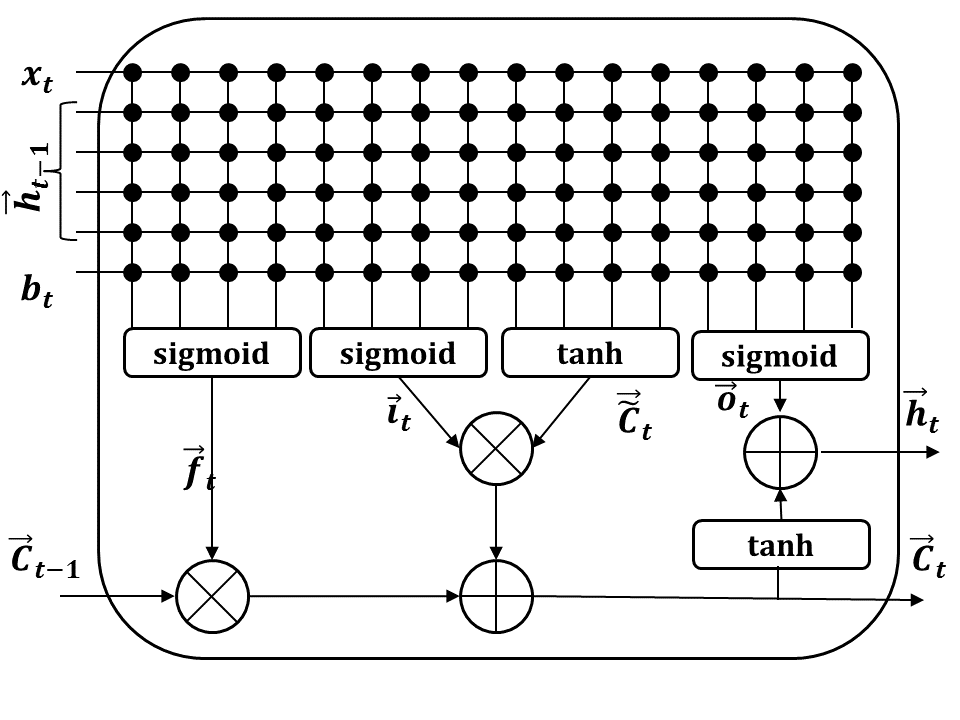}
    \caption{LSTM cell diagram}
    \label{celldiagram}
\end{figure}

A closer look into an LSTM cell structure can be seen in the Fig. \ref{celldiagram}. The input data of LSTM unit is a concatenated vector of new input data  $x_{t}$ and data from a previous cell $\bm{h_{t-1}}$. Biases $b_{t}$ are used to identify zero inputs. The concatenated vector is multiplied by a weight matrix and obtained outputs go through activation functions (either sigmoid or hyperbolic tangent) to form gate values. The output values of forget gate $\bm{f_{t}}$ after sigmoid layer is between 0 and 1. Hadamard multiplication of $\bm{f_{t}}$  with previous cell state $\bm{C_{t-1}}$ is used to decide whether to keep or partially/completely delete information on $\bm{C_{t-1}}$ in the current cell. Similarly the input gate output $\bm{i_{t}}$ contributes to the new cell state $\bm{C_{t}}$ by deciding to block or pass a new candidate cell state $\bm{\tilde{C}_{t}}$. Combination of some part of new and some part of old information then produces new candidate cell state $\bm{C_{t}}$ mathematically shown in (\ref{eqn4}). Eventually, output gate $\bm{o_{t}}$ decides how much of the filtered version of $\bm{C_{t}}$ forms a new cell output $\bm{h_{t}}$ as shown in (\ref{eqn6}). $\bm{C_{t}}$ is filtered through hyperbolic tangent function to have the outputs between -1 and 1. 

\par
Operation of LSTM cell can be described by following equations \cite{olah2015understanding}:

\begin{equation}\label{eqn1}
\bm{f_{t}}=\sigma(\bm{W_{f}x_{t}}+\bm{U_{f}h_{t-1}}+\bm{b_{f}})
\end{equation}
\begin{equation}\label{eqn2}
\bm{i_{t}}=\sigma(\bm{W_{i}x_{t}}+\bm{U_{i}h_{t-1}}+\bm{b_{i}})
\end{equation}
\begin{equation}\label{eqn3}
\bm{\tilde{C}}=tanh(\bm{W_{\tilde{C}}x_{t}}+\bm{U_{\tilde{C}}h_{t-1}}+\bm{b_{\tilde{C}}})
\end{equation}
\begin{equation}\label{eqn4}
\bm{C_{t}}=\bm{i_{t}}\bigodot \bm{\tilde{C}_{t}}+\bm{f_{t}}\bigodot \bm{C_{t-1}}
\end{equation}
\begin{equation}\label{eqn5}
\bm{o_{t}}=\sigma(\bm{W_{o}x_{t}}+\bm{U_{o}h_{t-1}}+\bm{b_{o}})
\end{equation}
\begin{equation}\label{eqn6}
\bm{h_{t}}=\bm{o_{t}}\bigodot \bm{tanh(C_{t}}),
\end{equation}

Since the number of hidden units in our LSTM cell is chosen to be four, the size of matrices  $\bm{U_{i}}$, $\bm{U_{f}}$, $\bm{U_{\tilde{C}}}$ and $\bm{U_{o}}$ are [4x4]. And sizes of matrices $\bm{W_{i}}$, $\bm{W_{f}}$, $\bm{W_{\tilde{C}}}$, $\bm{W_{o}}$, $\bm{b_{i}}$, $\bm{b_{f}}$, $\bm{b_{\tilde{C}}}$ and $\bm{b_{o}}$ are [1x4].  The resulting size of the matrix in a LSTM unit for the given problem is [6x16]. Weights and biases were constrained to the range [-1;1]. Upon training in Python Keras, their values were extracted to build the circuit.

\subsection{Vector-matrix multiplication circuit}

In hardware, weight matrix of LSTM cell can be implemented using memristor crossbar array \cite{smagulova2018memristor}, \cite{irmanova2017multi}. Memristor is a non-volatile element capable of remembering its resistance state. Typically it has two states $R_{on}$ and $R_{off}$ that can be controlled by applied voltage amplitude and pulse duration. Memristor existence was postulated by L.Chua back in 1971 \cite{chua1971memristor} and HP Labs announced its in 2008 \cite{williams2008we}, \cite{strukov2008missing}. It is a promising element due to nanoscale size, absence of leakage current and reprogramming ability. Using memristors in a crossbar array to perform vector-matrix multiplication benefits in fast computational speed and small area.

\par
\begin{figure}[h]
   \centering
   \includegraphics[scale=.35]{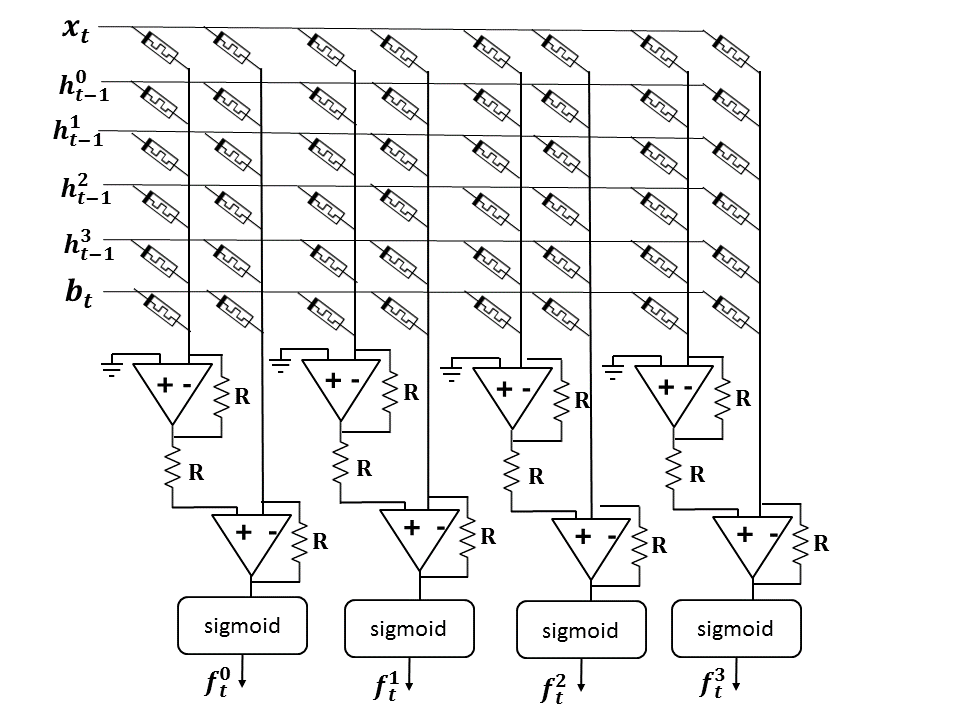}
    \caption{Vector-matrix multiplication circuit of the forget gate }
    \label{forgetgate}
\end{figure}

 
Since weight value in LSTM cell can take both positive and negative values, it can be represented as a difference of two memristor conductances \cite {hasan2017chip}. This doubles the number of memristors in a matrix. $R_{on}$ and $R_{off}$ of memristors were chosen to be 10k$\Omega$ and 10M$\Omega$, respectively.   Fig.\ref{forgetgate} shows implementation of forget gate of the LSTM cell. Similar approach can be used to construct the rest of the gates.
\par

The vector-matrix multiplication circuit input (input, hidden unit) values are the same as for the software implementation case between 0 and 1. Only exception is bias input values (not bias weights, though weights can be adjusted accordingly) in the circuit are higher by 0.3V in the LSTM layer and by 0.25V in the dense layer than that of bias input values obtained from the software implementation. It is done to make up for the stage-wise loss of voltage values that propagate through the circuit. The bias input value in LSTM layer is 1.5 and in the dense (feed-forward neural network) layer is 0.0239 (in software).

\subsection{Sigmoid and hyperbolic tangent function circuits}
\begin{figure}[h]
   \centering
   \includegraphics[scale=.38]{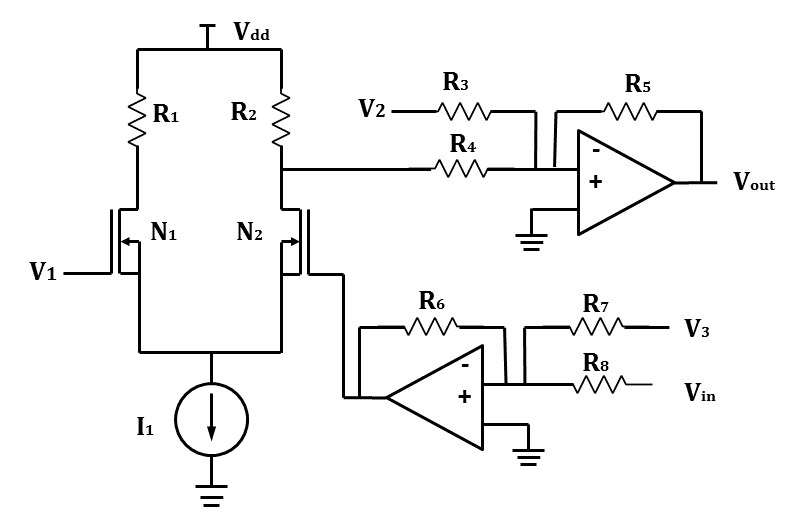}
    \caption{Activation function circuit}
    \label{activfunc}
\end{figure}

\begin{figure}[!hh]
\centering
	\subfigure[]{
    \includegraphics[width=0.215\textwidth]{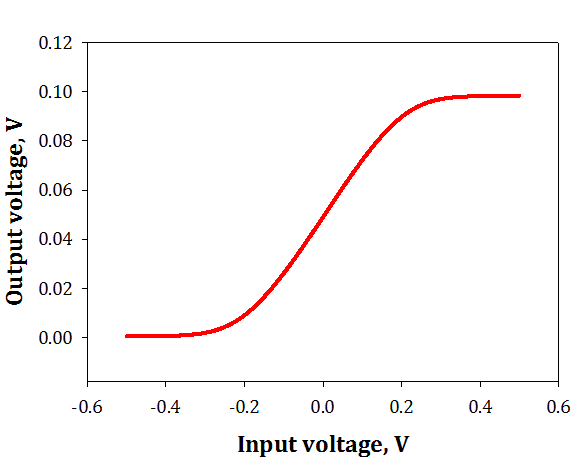}}
    \label{sigmout}
    \subfigure[]
   { \includegraphics[width=0.23\textwidth]{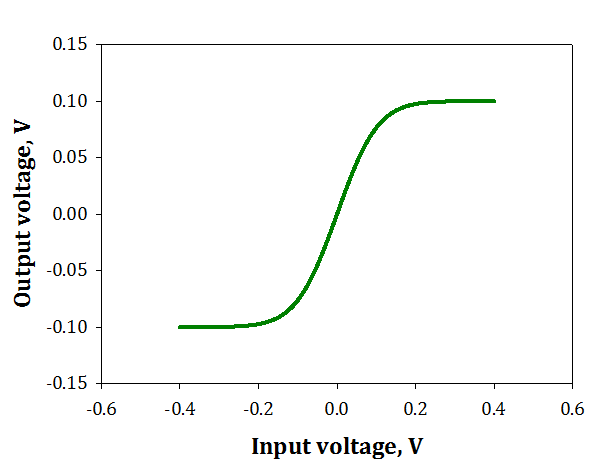} }
     \label{tanh}
     \caption{Activation function circuits' outputs multiplied by -1: a) sigmoid and b) hyperbolic tangent }
     \label{3T}
\end{figure}

Sigmoid and hyperbolic tangent functions can be obtained using circuit in Fig. \ref{activfunc}. It basically employs the property of differential amplifier -- gradual and smooth increase of the output voltage when the differential input is swept between a desired range. The desired output range and form can be obtained by varying supply voltage $V_{dd}$, current $I_1$, and the sizes of NMOS transistors ($N_1$ and $N_2$). Voltage source values of $V_1$, $V_2$, and $V_3$ are used to shift the output values to match the graphs of the sigmoid and hyperbolic tangent functions. Since these two functions are different, the above mentioned parameters also change for each function. DC transfer characteristics for sigmoid and hyperbolic tangent function circuits are shown in Fig.\ref{3T}$a$ and Fig.\ref{3T}$b$, respectively. The graphs' input and output ranges are scaled down by -10 (negative part is canceled at later stages) to meet the operation range of the other circuit elements.
\subsection{Four-quadrant multiplier circuit}

\begin{figure}[h]
   \centering
   \includegraphics[scale=.35]{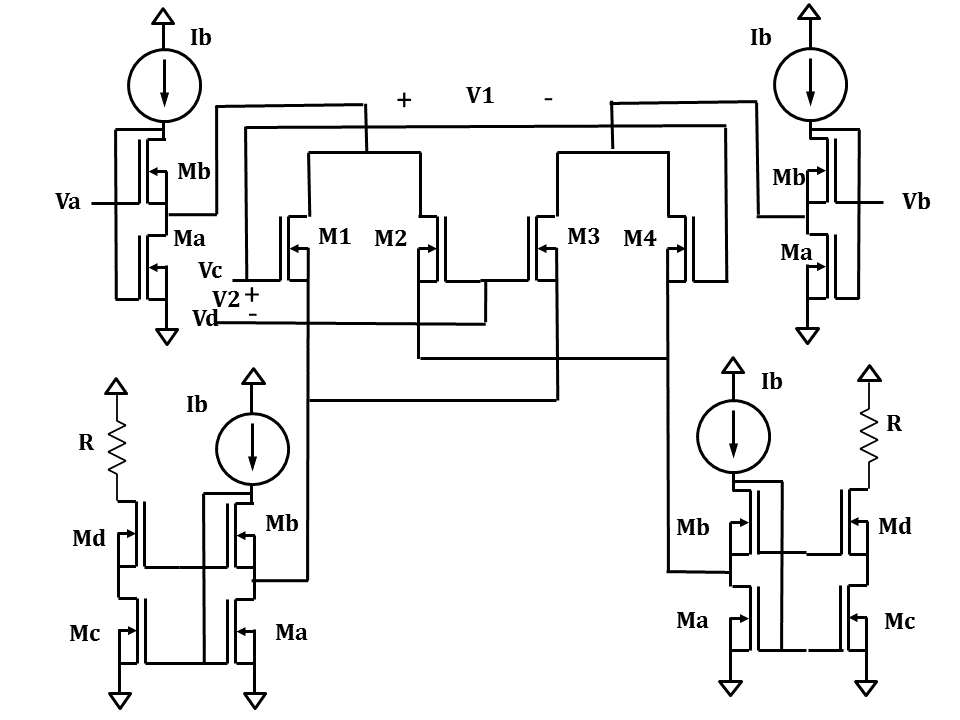}
    \caption{Four-quadrant multiplier circuit}
    \label{multiplier}
\end{figure}

\begin{figure}[h]
   \centering
   \includegraphics[scale=.45]{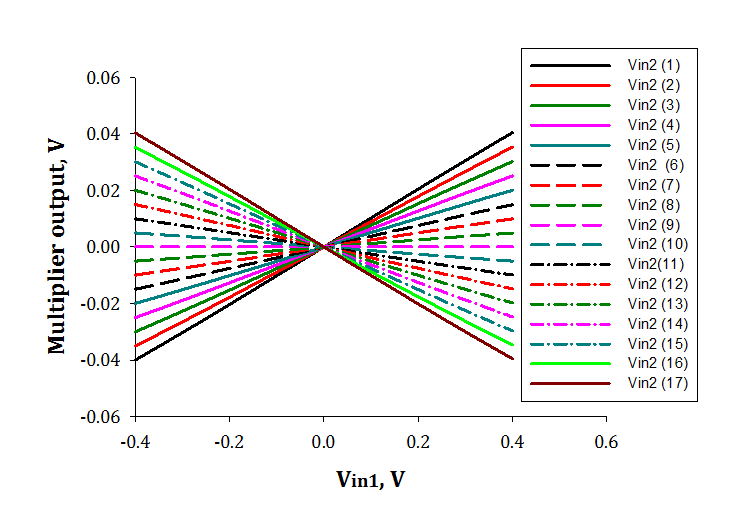}
    \caption{DC transfer characteristics of a four-quadrant voltage multiplier }
    \label{multout}
\end{figure}

Fig. \ref{multiplier} shows a low-voltage four quadrant transconductance CMOS multiplier from \cite{ramirez2004low}. Its core cell consists of NMOS transistors M1-M4. Current source $I_b$ and transistors $M_a$ and $M_b$ form a flipped voltage follower cell. It is characterized by low impedance for current sensing purposes. The multiplier circuit was further extended to have single-ended output. It has been done by scaling down the source voltages of transistors $M_a$, then amplifying their difference, and finally shifting plus amplifying the amplified difference to get highly accurate output value. In addition, current sources in Fig. \ref{multiplier} were replaced with its CMOS implementations. Voltage transfer characteristics of the multiplier circuit is shown Fig. \ref{multout}. The outputs are scaled down by -4 for the same reason as in the activation function circuits.  

\section{Simulation results}

\par
\begin{figure}[ht]
   \centering
   \includegraphics[scale=.40]{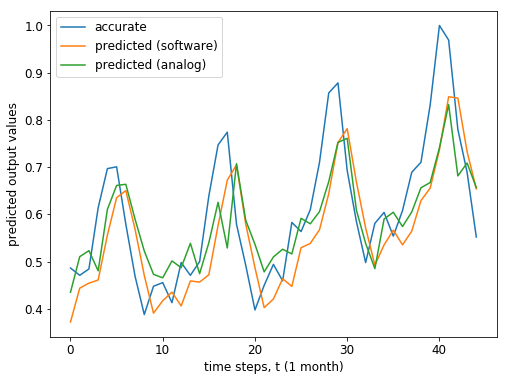}
    \caption{Graphical prediction comparison}
    \label{comparison}
\end{figure}

Simulation performance results obtained from software and hardware (analog) closely match as seen in Fig. \ref{comparison} and both of them repeat the general trend in the plot. Numerical comparison reveals that MSE and RMSE for the software prediction are 0.0112 and 0.1059, respectively. While, MSE and RMSE for the analog prediction are 0.0101 and 0.1004, respectively.

The total circuit simulation time for predicting 45 test points is 3.96 ms with each cycle taking 88 $\mu$s. First 40 microseconds are spent for the computation of $h_{t-1}$ and $C_{t-1}$ for all four units with 10 $\mu$s being a sub-cycle of computing outputs of a single hidden unit. In fact, 2 microseconds of those 10 microseconds are used as an intentional delay to avoid convergence issues. Sub-cycles are realized using pass-logic circuits and corresponding control signals. At the end of each sub-cycle time, the outputs are stored at memory circuits consisting of op-amp buffer and a capacitor. They are to be used in the second time step calculations. After the first time step ends, intentional 2 $\mu$s delay is inserted. The next 40  $\mu$s starting from 42  $\mu$s to 82  $\mu$s are used for the computation of $h_{t}$ and $C_{t}$ for all hidden units in the same sub-cycle manner. Again after the second time step, 2  $\mu$s delay is introduced. Starting from 84  $\mu$s to 87  $\mu$s, the circuit computes the final output value - time series prediction value. Final stage is given short time interval of 3  $\mu$s, because the circuit computes only the weighted sum of $\bm{h_{t}}$ through a dense layer and adds a bias to it. Lastly, 1  $\mu$s intentional delay is introduced and the first cycle ends.

Total maximum power consumption for a single LSTM cell is 210.67 mW when all the inputs ($x_{t}$ and $\bm{h_{t-1}}$) are set to 1V and the bias input $b_t$ is set to 1.8V; and its area is 58569$\mu$$m^2$.



\section{Conclusion}
\label {conclusion}

This paper proposed voltage-based LSTM circuit design for predicting the number of international airplane passengers. Having voltage-based LSTM circuit reduces the overhead of converting between currents and voltages which significantly reduces circuit area and complexity. In addition, it makes it easy to adjust the intermediate node voltages to desired voltage values during the stage-wise building of the circuit.
From obtained simulation results, it is clear that the prediction results closely match between the hardware and software implementations. 
Simulation times can be further reduced by adjusting the control voltage signals by making sure that there are no convergence issues.
\bibliographystyle{IEEEtran}
\bibliography{bibliography}

\end{document}